%% file: tsp.tex
\documentclass[camera]{jpaper}

\usepackage[nocompress]{cite}
\usepackage{algorithmic}
\usepackage{array}

\makeatletter
\let\MYcaption\@makecaption
\makeatother

\usepackage[font=footnotesize]{subcaption}

\makeatletter
\let\@makecaption\MYcaption
\makeatother

\usepackage{fixltx2e}
\usepackage{dblfloatfix}
\usepackage[nolessnomore, italic]{mathastext}
\usepackage[T1]{fontenc}
\usepackage[usenames,dvipsnames,svgnames,table]{xcolor}
\usepackage{multirow}
\usepackage{hhline}
\usepackage[normalem]{ulem}
\usepackage{enumitem}
\usepackage{setspace}
\usepackage{indentfirst}
\usepackage{footmisc}
\usepackage{fancyhdr}
\usepackage{authblk}
\usepackage[us,12hr]{datetime}
\usepackage[keeplastbox]{flushend}
\usepackage[hidelinks]{hyperref}

\newif\ifcameraready
\camerareadytrue

\newcommand{\versionnum}[0]{2.1}

\ifcameraready
\else
\fi

\newcommand{\paratitle}[1]{\vspace{6pt}\textbf{#1.}}
\newcommand{\incircle}[1]{\raisebox{0.5pt}{\protect \textcircled{\raisebox{-0.8pt}{#1}}}}

\ifcameraready
  \newcommand{\todo}[1][]{}
  \newcommand{\chVII}[1]{#1}
\else
  \newcommand{\todo}[1][]{\textbf{\fcolorbox{black}{red}{\color{white}{TODO}}} \underline{$\overline{\hbox{\emph{#1}}}$}}
  \newcommand{\chVII}[1]{{\color{BrickRed}#1}}
\fi
\newcommand{\changes}[1]{{#1}}
\newcommand{\oldchangesI}[1]{{#1}}
\newcommand{\oldchangesII}[1]{{#1}}
\newcommand{\oldchangesIII}[1]{{#1}}
\newcommand{\oldchangesIV}[1]{{#1}}

\ifcameraready

\else

\fi

\begin{document}
%
\title{Characterizing, Exploiting, and Mitigating\\Vulnerabilities in MLC NAND Flash Memory Programming}


\author{
  {Yu Cai$^{1}$\qquad}
  Saugata Ghose$^{2}$\qquad
  Yixin Luo$^{1,2}$%
\vspace{2pt}\\
  Ken Mai$^{2}$\qquad
  Onur Mutlu$^{3,2}$\qquad
  Erich F. Haratsch$^{1}$}
\affil{{\em $^{1}$Seagate Technology\qquad
  $^{2}$Carnegie Mellon University\qquad
  $^{3}$ETH Z{\"u}rich}%
}


%


\maketitle




%

\input{sections/abstract}
\input{sections/summary}
\input{sections/related}
\input{sections/impact}
\input{sections/conclusion}

\section*{\chVII{Acknowledgments}}

\chVII{We thank the anonymous reviewers for their
feedback on our HPCA 2017 paper~\cite{cai.hpca17}. 
This work is partially supported by the Intel Science
and Technology Center, CMU Data Storage Systems Center,
and NSF grants 1212962 and 1320531.}



%


{
\bibliographystyle{IEEEtranS}
\bibliography{refs}
}

\end{document}

%% file: sections/abstract.tex

\begin{abstract}

This paper summarizes our work on experimentally analyzing, exploiting,
and addressing vulnerabilities in
multi-level cell NAND flash memory \chVII{programming}, which was published in the industrial 
session of HPCA 2017~\cite{cai.hpca17},
and examines the work's significance and future potential.
Modern NAND flash memory chips use \emph{multi-level cells} (MLC), which
store two bits of data in each cell, to improve chip density.
As MLC NAND flash memory scaled down to smaller manufacturing process
technologies, manufacturers adopted a \emph{two-step programming method} to
improve reliability.  In two-step programming, the two bits of a multi-level 
cell are programmed using two separate steps, in order to \chVII{minimize} the amount of
cell-to-cell program interference induced on neighboring flash cells.

In this work, we demonstrate that two-step programming exposes 
new reliability and security vulnerabilities in state-of-the-art MLC
NAND flash memory.
We experimentally characterize contemporary 1X-nm (i.e., 15--19nm)
flash memory chips, and
find that a \emph{partially-programmed} flash cell
(i.e., a cell where the second programming step has not yet been performed) is
much more vulnerable to cell-to-cell interference and read disturb than
a fully-programmed cell.
We show that it is possible to exploit 
these vulnerabilities on solid-state drives (SSDs) to alter the 
partially-programmed data, causing (potentially malicious) data corruption. 
Based on our observations, we propose several new mechanisms
that eliminate or mitigate these vulnerabilities in partially-programmed cells, 
and at the same time increase flash memory lifetime by 16\%.

\end{abstract}

%% file: sections/summary.tex

\section{Introduction}
\label{sec:motivation}

Solid-state drives (SSDs), which consist of NAND flash memory chips, are
widely used for storage today due to significant decreases in the per-bit cost of NAND flash memory,
which, in turn, have driven great increases in SSD capacity. These 
improvements have been enabled by both aggressive process technology scaling and the 
development of \emph{multi-level cell} (MLC) technology. 
NAND flash memory stores data by changing the threshold voltage of each
flash cell, where a cell consists of a \emph{floating-gate transistor}~\cite{kahng.bell67, momodori.patent88, masuoka.iedm87}.
In single-level cell (SLC) flash memory, the threshold voltage range could represent only a single bit of data.
A multi-level cell uses the same
threshold voltage range to represent \emph{two} bits of data within a single
cell (i.e., the range is split up into four windows, known as \emph{states}, where each state
represents one of the data values 00, 01, 10, or 11),
thereby doubling storage capacity~\cite{cernea.jssc09,hara.isscc05,lee.jssc11,park.jssc08,suh.jssc95,cai.date12}.
In a NAND flash memory chip, a row of cells is connected together by a common
\emph{wordline}, which typically spans 32K--64K cells.
Each wordline contains two \emph{pages} of data, where a page is the granularity
at which the data is read and written (i.e., programmed).
The most significant bits (MSBs) of all cells on the same wordline are
combined to form an \emph{MSB page}, and the least significant bits (LSBs) of all cells on
the wordline are combined to form an \emph{LSB page}~\cite{cai.iccd13}.

To precisely control the threshold voltage of a flash cell, the flash memory device
uses \emph{incremental step pulse programming} 
(ISPP)~\cite{cernea.jssc09,hara.isscc05,lee.jssc11,suh.jssc95}.  ISPP
applies multiple short pulses of a high programming voltage to each cell
in the wordline being programmed, with each pulse 
increasing the threshold voltage of the cell by some small amount.
SLC and older MLC devices programmed the threshold voltage in \emph{one 
shot}, issuing all of the pulses back-to-back to program \emph{both} bits of data 
at the same time. However, as flash memory scales down \chVII{to smaller technology nodes}, the distance 
between neighboring flash cells decreases, which in turn increases the 
\emph{program interference} that occurs due to cell-to-cell coupling. This 
program interference causes errors to be introduced into neighboring cells 
during programming~\cite{cai.sigmetrics14,dong.tcas10,lee.edl02,lee.icassp12,park.jssc08,cai.iccd13}. 
To reduce this interference by half~\cite{cai.iccd13}, 
manufacturers have been using \emph{two-step programming} for MLC NAND flash
memory since the 40nm technology node~\cite{park.jssc08}.
A large fraction of SSDs on the market today use sub-40nm MLC NAND flash memory.

\begin{figure}[h]
  \centering
  \includegraphics[width=0.875\columnwidth, trim=57 194 42 167, clip]{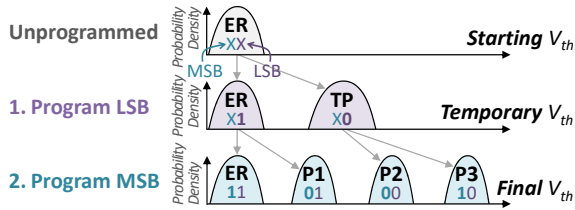}%
  \caption{Starting (after erase), temporary (after LSB programming), and final (after MSB programming) states for two-step programming.
  Reproduced from \cite{cai.hpca17}.}%
  \label{fig:twostep}%
\end{figure}

Two-step programming stores each bit within an MLC \chVII{flash memory cell}
using two \emph{separate, partial programming} steps,
as shown in Figure~\ref{fig:twostep}.
An unprogrammed cell starts in the erased (ER) state.
The first programming step programs the LSB page:
for each flash cell within the page, the cell is
\emph{partially programmed} depending on the LSB being written to the cell.
If the LSB of the cell 
should be 0, the cell is programmed into a temporary program state (TP); 
otherwise, it remains in the ER state.
The maximum voltage of a partially-programmed cell
is approximately half of the maximum possible threshold voltage of 
\oldchangesIII{a fully-programmed flash cell}. 
In \oldchangesII{its} second step, two-step programming programs the MSB page:
it reads the LSB value into a buffer inside the flash \oldchangesIII{chip} \oldchangesII{(called the \oldchangesIII{\emph{internal LSB buffer}})} to determine
the partially-programmed state of the cell's threshold voltage, and then partially 
programs the cell again, 
depending on whether the MSB of the cell is a 0 or a 1.
The second programming step moves the threshold voltage 
from \oldchangesII{the} partially-programmed state to \oldchangesII{the desired} final state
(i.e., ER, P1, P2, or P3).
By breaking MLC programming into two separate steps, manufacturers
\emph{halve} the program interference of each programming operation~\cite{lee.edl02, cai.iccd13}.
\oldchangesIII{The SSD controller employs \emph{shadow program
sequencing}~\chVII{\cite{cai.procieee17, cai.procieee.arxiv17, cai.bookchapter.arxiv17,
choi.presentation10,cai.iccd13,park.dac16}},
which interleaves the partial programming steps of a cell 
with} the partial programming \oldchangesIII{steps} of neighboring 
cells to ensure that a \emph{fully-programmed}
cell experiences interference only from a single neighboring partial programming 
step.\footnote{We refer the reader to our prior works~\cite{cai.procieee17, cai.procieee.arxiv17,
cai.date13, cai.date12, cai.hpca15, cai.dsn15, luo.jsac16, cai.sigmetrics14, cai.iccd12, cai.itj13,
cai.hpca17, cai.iccd12, cai.bookchapter.arxiv17} for a detailed background on 
NAND flash memory. \chVII{Our recent survey paper~\cite{cai.procieee17,
cai.procieee.arxiv17, cai.bookchapter.arxiv17} provides an extensive survey
of the state-of-the-art in NAND flash memory.}}

\section{Error Sources in Two-Step Programming}
\label{sec:errors}

\begin{figure}[h]
  \centering
  \includegraphics[width=\linewidth, trim=38 248 38 150, clip]{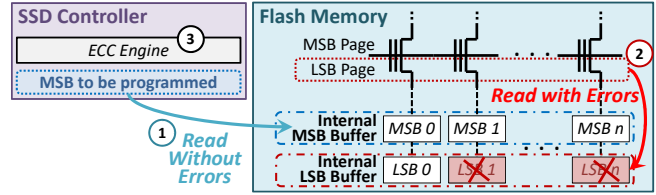}%
  \caption{\oldchangesII{In the second step of
  two-step programming, LSB data does not go to the controller, and is not corrected when read into 
  the internal LSB buffer, resulting in program errors}. Reproduced from \cite{cai.hpca17}.}%
  \label{fig:readerrors}%
\end{figure}

In our HPCA 2017 paper~\cite{cai.hpca17}, we demonstrate that two-step programming introduces \emph{new 
\oldchangesII{possibilities}} for \oldchangesI{flash memory} errors that \oldchangesI{can corrupt} some of 
the data \oldchangesI{stored within flash cells} without accessing them, and that these errors
can be exploited to design malicious attacks. As there is a delay between programming 
the LSB and the MSB of a single cell due to the \oldchangesI{interleaved} writes to 
neighboring cells, raw bit errors can be introduced into the already-programmed LSB 
page \oldchangesI{\emph{before}} the MSB page is programmed. 
These errors can cause a cell to be programmed
to an incorrect state in the second programming step.
During the \oldchangesII{second} step, both the 
MSB and LSB \oldchangesII{of each cell} are required to determine the 
final target threshold voltage \oldchangesII{of the cell.}  As shown in 
Figure~\ref{fig:readerrors}, the \oldchangesI{data to be programmed into 
the \oldchangesII{MSB}} is loaded from the \oldchangesII{SSD} controller to \oldchangesII{the internal 
MSB buffer (\incircle{1} in the figure)}. Concurrently, the \oldchangesII{LSB data} is loaded \changes{into the}
\oldchangesII{internal LSB buffer} from the \oldchangesI{flash memory wordline \oldchangesII{(\incircle{2})}.
\oldchangesI{By \oldchangesII{buffering the LSB data} inside the flash chip and not in the
\oldchangesII{SSD} controller, flash manufacturers avoid data transfer between the chip and the
controller \oldchangesII{during the second programming step, thereby reducing the step's} latency.  Unfortunately, this means that
the errors loaded from the \oldchangesII{internal LSB buffer \emph{cannot}} be corrected 
as they would otherwise
be during a read operation, because the error correction (ECC) engine resides \oldchangesII{only \emph{inside
the controller} (\incircle{3}), and not inside the flash chip}.}}
As a result, the final cell voltage can be \oldchangesI{\emph{incorrectly}} set 
during MSB programming, \emph{permanently corrupting} the LSB data.

We briefly discuss two sources of errors that can corrupt LSB data, \oldchangesII{and characterize} 
their impact on \emph{real} state-of-the-art 1X-nm (i.e., 15-19nm) MLC NAND flash 
chips. 
We perform our characterization using an FPGA-based flash testing 
platform~\cite{cai.fccm11, cai.date12} that allows us to issue commands directly 
to raw NAND flash memory chips. In order to determine the threshold voltage 
stored within each cell, we use the \emph{read-retry} mechanism built into 
modern SSD controllers~\cite{cai.date13,cai.iccd13,zuolo.tcad15,shim.vlsit11}. 
Throughout this work, we present \emph{normalized} voltage values, as actual 
voltage values are proprietary information to flash manufacturers.
Our complete characterization results can be found in our HPCA 2017
paper~\cite{cai.hpca17}.

\subsection{Cell-to-Cell Program Interference}
\label{sec:errors:celltocell}

\begin{figure}[h]
  \centering
  \includegraphics[width=.8\linewidth, trim=153 181 109 125, clip]
{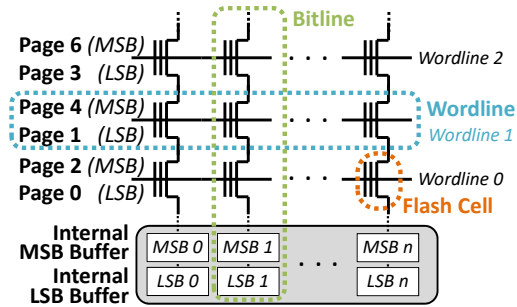}%
  \caption{\chVII{Internal architecture of a block of all-bit-line (ABL) flash
 memory.  Reproduced from \cite{cai.hpca17}.}}%
  \label{fig:ablflash}%
\end{figure}

The first error source, \emph{cell-to-cell program interference}, 
introduces errors into a flash cell when neighboring cells are programmed, 
as a result of parasitic capacitance coupling~\cite{dong.tcas10,lee.edl02,cai.iccd13, 
cai.sigmetrics14, grupp.micro09, cooke.fms07, cai.procieee17, cai.procieee.arxiv17,
cai.bookchapter.arxiv17}. 
While two-step programming reduces program interference for fully-programmed
cells, we find that interference during two-step programming is a significant error source for
\emph{partially-programmed cells}.  
\chVII{As an example, we look at a flash block in the 
 commonly-used all-bit-line (ABL) flash architecture~\cite{cernea.jssc09, cernea.isscc08,
 cai.iccd13}, which is shown in Figure~\ref{fig:ablflash}.
 After} the LSB page on Wordline~1 (Page~1 \chVII{in Figure~\ref{fig:ablflash}}) is
 programmed, the next two pages that are
programmed (Pages~2 and 3) reside on \oldchangesI{directly-adjacent} wordlines. 
Therefore, before the MSB page on Wordline~1 
\oldchangesII{(Page~4)} is programmed,
the LSB page (Page~1)
could be \emph{susceptible} to program interference
when Pages~2 and 3 are programmed.

\begin{figure}[h]
  \centering
  \includegraphics[width=0.40\columnwidth, trim=305 291 360 215, clip]{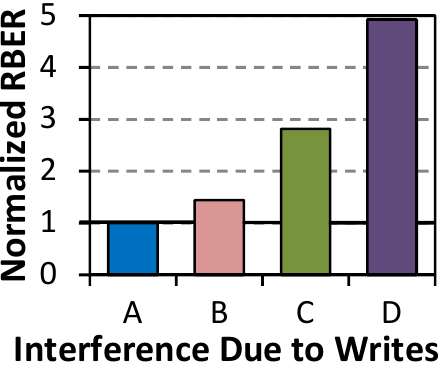}%
  \quad%
  \includegraphics[width=0.27\linewidth, trim=350 247 350 255, clip]{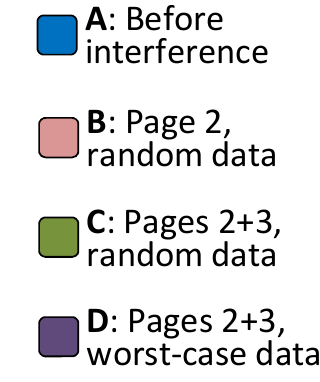}%
  \caption{Normalized raw bit error rate of partially-programmed Page~1, before and after cell-to-cell program interference.
  Adapted from \cite{cai.hpca17}.}%
  \label{fig:partialBER}%
\end{figure}

Figure~\ref{fig:partialBER} shows the
measured raw bit error rate for Page~1 in real NAND flash
memory devices after 
four different times, normalized to the error rate just after Page~1 is
programmed: 
\begin{enumerate}[label=\Alph*.]
  \item Just after Page~1 is programmed (no interference),
  \item Page~2 is programmed with pseudo-random data,
  \item Pages~2 and 3 are programmed with pseudo-random data,
  \item Pages~2 and 3 are programmed with \oldchangesIII{a data pattern that
    induces the \emph{worst-case} program interference}.
\end{enumerate}
We observe that the amount of interference is
especially high when Pages~2 and 3 \chVII{in Figure~\ref{fig:ablflash}} are written with the worst-case data pattern,
after which the raw bit error rate of Page~1 is \emph{4.9x the rate before interference}.
\chVII{Note that the worst-case data pattern that we write to Pages~2
and 3 \emph{requires no knowledge of the data stored within Page~1}~\cite{cai.hpca17}.}

\subsection{Read Disturb}
\label{sec:errors:readdisturb}

The second error source, \emph{read disturb}, disrupts the 
contents of a flash cell when another cell is read~\chVII{\cite{cai.dsn15,ha.apsys13,
cooke.fms07,grupp.micro09,mielke.irps08,takeuchi.jssc99,papandreou.glsvlsi14,
cai.procieee17, cai.procieee.arxiv17, cai.bookchapter.arxiv17}}. 
NAND flash memory cells are organized into multiple \emph{flash blocks}
(two-dimensional cell arrays), where each block contains a set of \emph{bitlines} that connect 
multiple flash cells in series. To accurately read the value from 
one cell, 
the SSD controller applies a \emph{pass-through voltage} to turn on the
\emph{unread} cells on the bitline, which allows the value to propagate through the bitline.
Unfortunately, this pass-through 
voltage induces \oldchangesII{a} \emph{weak programming \oldchangesII{effect}} \oldchangesII{on \oldchangesIII{an} unread cell: it slightly increases}
the cell threshold voltage\oldchangesI{~\cite{cai.dsn15, cai.procieee17, cai.procieee.arxiv17, cai.bookchapter.arxiv17}}. 
As more neighboring cells within a block are read, 
\oldchangesII{\changes{an unread} cell's threshold voltage can increase
enough \oldchangesIII{to change}} \oldchangesII{the data value stored in the 
cell}~\cite{cai.dsn15,ha.apsys13, cai.procieee17, cai.procieee.arxiv17, cai.bookchapter.arxiv17,papandreou.glsvlsi14}. 
\oldchangesI{In two-step programming, \oldchangesII{a partially-programmed cell is more
likely to have a lower threshold voltage than a fully-programmed cell,
and the weak programming effect is stronger on cells with a lower threshold voltage.}
Measuring errors in real NAND flash memory devices, we find that the raw bit 
error rates for an LSB page in a partially-programmed or unprogrammed wordline is
\emph{an order of magnitude greater} than the rate for an LSB page in a fully-programmed
wordline.}
However, existing read 
disturb management solutions are designed to protect fully-programmed 
cells~\cite{cai.dsn15,frost.patent10,schushan.patent14, ha.apsys13, ha.tcad16, kim.patent12}, and offer little mitigation for 
\oldchangesI{partially-programmed \oldchangesII{cells}}.

\section{Exploiting Two-Step Programming Errors}
\label{sec:exploits}

Two major issues arise from the program interference and read disturb
vulnerabilities of partially-programmed and unprogrammed cells.
First, the vulnerabilities induce a large number of errors on these cells, 
exhausting the SSD's error correction capacity and limiting the SSD lifetime.
Second, the vulnerabilities can potentially allow \oldchangesIII{(malicious)} applications to
aggressively corrupt and change data belonging to other programs and
further hurt the SSD lifetime.  We present two example sketches of potential
exploits in our HPCA 2017 paper~\cite{cai.hpca17}, which we briefly summarize 
here.

\subsection{Sketch of Program Interference Based Exploit}
\label{sec:exploits:celltocell}

\begin{figure}[h]
  \centering
  \includegraphics[width=\linewidth, trim=41 141 38 115, clip]{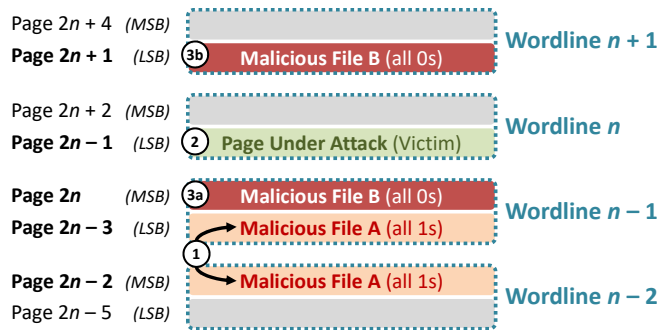}%
  \caption{Layout of data within a flash block during a program interference based exploit.
  Reproduced from \cite{cai.hpca17}.}%
  \label{fig:interferencelayout}%
\end{figure}

In this exploit, a malicious application 
can induce a significant amount of \emph{program interference} onto a flash page that 
belongs to another, benign victim application, corrupting the page and 
shortening the \changes{SSD} lifetime. 
Recall from Section~\ref{sec:errors:celltocell} that writing the worst-case data pattern
can induce \oldchangesII{4.9x} the number of errors into a neighboring page (with respect to 
\changes{an} interference-free page).
The goal of this exploit is for a 
malicious application to write this worst-case \chVII{data} pattern in a way that
ensures
that the page \oldchangesIII{that is} disrupted belongs to the victim application, and that 
\oldchangesIII{the page that is disrupted}
experiences the greatest amount of program interference 
possible.
Figure~\ref{fig:interferencelayout} illustrates the contents of the pages within
neighboring 8KB \emph{wordlines} (rows of flash cells within a block).
The SSD controller uses \emph{shadow program sequencing} to interleave
partial programming steps to pages in ascending order of the page numbers shown 
on the left side of the figure.
A malicious application can write a small 16KB file with all 1s to prepare for the
attack (\incircle{1} in the figure), and then waits for the victim application to write 
its data to Wordline~$n$
(\incircle{2}). Once the victim writes its data, the malicious application then
writes all 0s to a second 16KB file (\incircle{3a} and \incircle{3b}).  
This induces the largest possible change in
voltage on the victim data, and can be used to flip bits within the data.
In our HPCA 2017 paper~\cite{cai.hpca17}, we discuss how a malicious application can
(1)~work around SSD scrambling and (2)~monitor victim application data writes.

\subsection{Sketch of Read Disturb Based Exploit}
\label{sec:exploits:readdisturb}

In this exploit, a malicious application
can induce a significant amount of \emph{read disturb} onto \emph{several} 
flash pages that belong to other, benign victim applications. 
Recall from Section~\ref{sec:errors:readdisturb} that the error rate
after read disturb for an LSB page in a partially-programmed wordline is an order of
magnitude greater than the error rate for an LSB page in a fully-programmed
wordline. The goal of this exploit is for a 
malicious application to quickly perform a large number of read operations in
a very short amount of time, to induce read disturb errors that corrupt both
pages already written to partially-programmed wordlines and pages that
\emph{have yet to be written}.
The malicious application
writes an 8KB file, with arbitrary data, to the SSD.  
Immediately after the file is
written, the malicious application repeatedly forces
\oldchangesII{the file system to send a new read request to the} 
SSD.
Each request induces read disturb on
the other wordlines within the flash block, causing the cell threshold voltages 
of these wordlines to increase.
After the malicious application finishes performing the repeated read
requests, a victim application writes data to a file.
As the SSD is unaware that an attack took place, it does not detect that the
data cannot be written correctly due to the increased cell threshold
voltages.  As a result, bit flips \oldchangesIII{can} occur in the victim
application's data.
Unlike the program interference exploit, \oldchangesIII{which attacks a 
single page,}
the read disturb exploit can corrupt \oldchangesIII{multiple pages with a single
attack},
and the corruption can affect pages written at a 
much later time \oldchangesII{than the attack} if the host write rate is low.

\section{Protection and Mitigation Mechanisms}
\label{sec:mech}

\begin{figure}[h]
  \centering
  \includegraphics[width=0.99\linewidth, trim=33 189 33 173, clip]{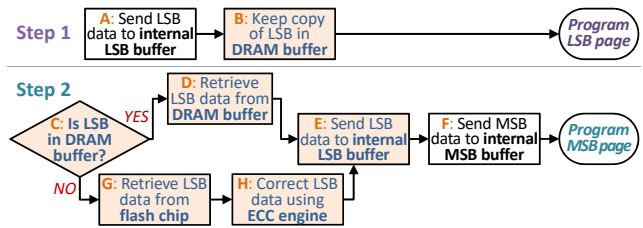}%
  \caption{\oldchangesII{Modified two-step programming}, using a DRAM 
  buffer for \oldchangesII{LSB data (\oldchangesIII{modifications shown in shaded boxes})}.
  Reproduced from \cite{cai.hpca17}.}%
  \label{fig:twostepbuffer}%
\end{figure}

We propose \oldchangesII{three} mechanisms to eliminate or mitigate \oldchangesII{the 
\oldchangesIII{program interference and read disturb vulnerabilities of partially-programmed and unprogrammed cells} due to two-step programming.
Table~\ref{tbl:mechanisms} 
summarizes the cost \oldchangesII{and benefits} of each mechanism.
We briefly discuss our three mechanisms here, and provide \oldchangesIV{more detail} on them
in our HPCA 2017 paper~\cite{cai.hpca17}.

\begin{table}[h]
  \centering
  \footnotesize
  \setlength{\tabcolsep}{0.45em}
  \renewcommand{\arraystretch}{0.95}
  \caption{\oldchangesII{Summary of our proposed protection mechanisms.  Reproduced from \cite{cai.hpca17}.\vspace{-5pt}}}%
  \label{tbl:mechanisms}%
  \begin{tabular}{|c||c|c|c|c|}
    \hline
    \multirow{2}{*}{\textbf{Mechanism}} & \textbf{Protects} & \multirow{2}{*}{\textbf{Overhead}} & \textbf{Error Rate} \\
    & \textbf{Against} & & \textbf{Reduction} \\
    \hhline{|=#=|=|=|}
    \emph{Buffering LSB Data} & interference & 2MB storage & \multirow{2}{*}{100\%} \\
    \emph{in the Controller} & read disturb & \oldchangesIII{1.3--15.7\%} latency & \\
    \hline
    \emph{Adaptive LSB Read} & interference & 64B storage & \multirow{2}{*}{\oldchangesIII{21--33\%}} \\
    \emph{Reference Voltage} & read disturb & 0.0\% latency & \\
    \hline
    \emph{Multiple Pass-Through} & \multirow{2}{*}{read disturb} & 0B storage & \multirow{2}{*}{72\%} \\
    \emph{Voltages} & & 0.0\% latency & \\
    \hline
  \end{tabular}%
\end{table}

Our first mechanism buffers LSB data in the SSD controller, eliminating
the need to read the LSB page from flash memory
at the beginning of the second programming step\oldchangesIII{, thereby \emph{completely 
eliminating the vulnerabilities}}.  
It} maintains a copy of all 
partially-programmed LSB data within \oldchangesIII{DRAM buffers that exist in the 
SSD near the controller.  Doing so ensures that the LSB data is read without 
any errors from the DRAM buffer, where it is \oldchangesIII{free from} the vulnerabilities
(instead of from the flash memory, where it incurs errors \chVII{that are not corrected}), in}
\oldchangesII{the second programming step.}
Figure~\ref{fig:twostepbuffer} shows a flowchart of our modified two-step programming
algorithm.
This solution increases the programming latency of the flash \oldchangesII{memory} by 
\oldchangesIII{4.9\% in the common case}, due to the \oldchangesI{long latency of sending 
\oldchangesII{the LSB data from the controller to \oldchangesIII{the internal} LSB buffer inside
flash memory}}.

\begin{figure}[h]
  \centering
  \begin{subfigure}[b]{0.365\columnwidth}%
    \includegraphics[width=\linewidth, trim=183 190 335 203, clip]{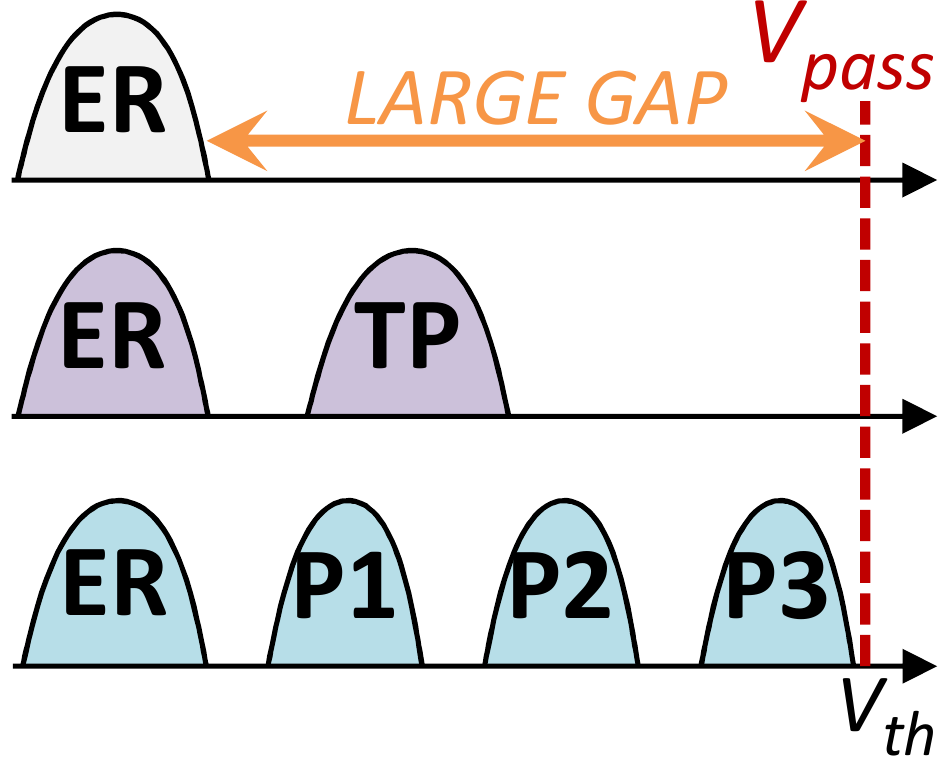}%
    \vspace{-10pt}%
	\caption{}%
    \label{fig:singlevpass}%
  \end{subfigure}%
  ~%
  \includegraphics[width=.260\columnwidth, trim=450 165 140 203, clip]{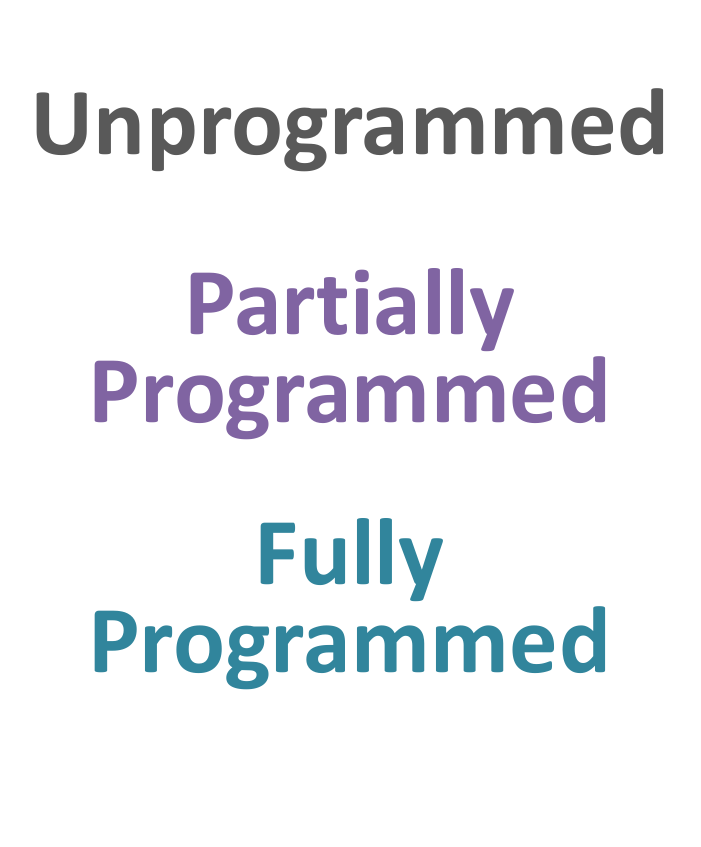}%
  ~%
  \begin{subfigure}[b]{0.365\columnwidth}%
    \includegraphics[width=\linewidth, trim=183 190 335 203, clip]{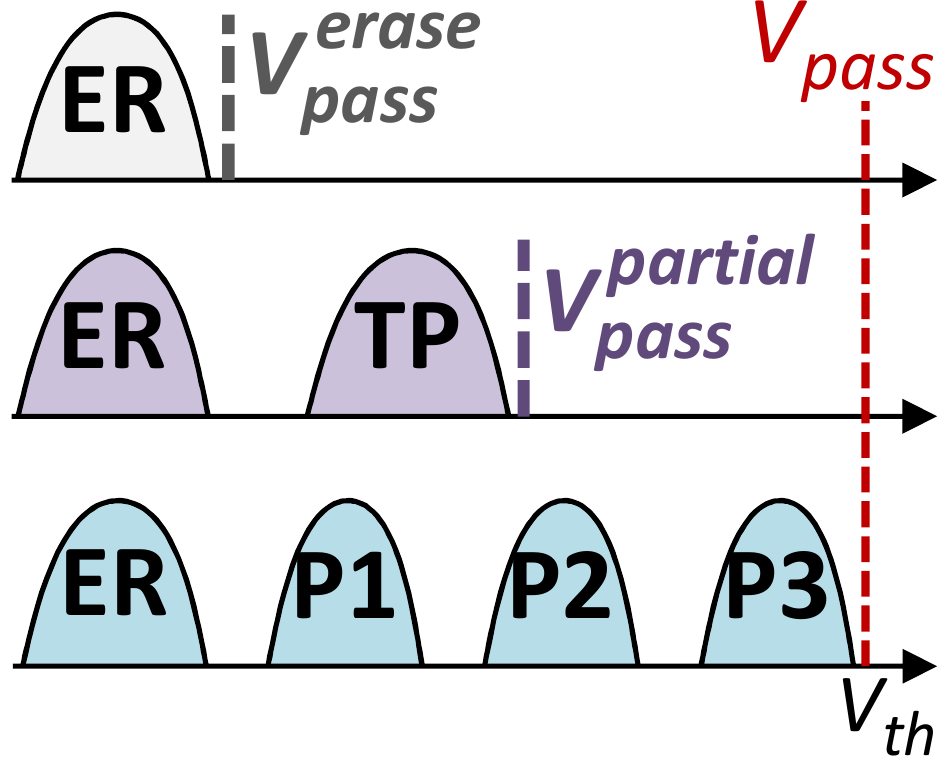}%
    \vspace{-10pt}%
	\caption{}%
    \label{fig:multiplevpass}%
    \label{fig:vpassoverview}%
  \end{subfigure}%
  \caption{(a) Applying \oldchangesII{single} $\boldsymbol{V_{pass}}$ to all unread wordlines; (b) Our multiple \oldchangesII{pass-through voltage} mechanism, \oldchangesII{where different voltages are applied based on the the wordline's programming status, to minimize the effects of read disturb}.
Reproduced from \cite{cai.hpca17}.}%
\end{figure}

The two other mechanisms that we develop largely mitigate (but do 
not \emph{fully} eliminate) the probability of two-step programming 
errors at much lower latency \oldchangesI{impact}. 
\oldchangesII{Our second mechanism \oldchangesIII{adapts the LSB read operation
to account for} threshold voltage changes
induced by program interference and read disturb.  It adaptively learns}
\oldchangesI{an \emph{optimized}}
read reference voltage for \oldchangesIII{LSB} data, lowering the 
probability of an LSB read error. 
\oldchangesII{Our third} mechanism greatly reduces the 
\oldchangesII{errors induced during read disturb,} by customizing the pass-through voltage for 
unprogrammed and partially-programmed flash cells. 
State-of-the-art SSDs apply a single pass-through voltage ($V_{pass}$) to all
of the unread cells, as shown in Figure~\ref{fig:singlevpass}.
This leaves a large gap between the pass-through voltage and the threshold 
voltage of a partially-programmed or unprogrammed cell, which greatly increases
the impact of read disturb~\cite{cai.dsn15, cai.hpca17}.
To minimize this gap, and, thus, the impact of read disturb, we propose to use
\emph{three} pass-through voltages, as shown in Figure~\ref{fig:multiplevpass}:
$V_{pass}^{erase}$ for unprogrammed cells, 
$V_{pass}^{partial}$ for partially-programmed cells,
and the same pass-through voltage as before ($V_{pass}$) for
fully-programmed cells.
\oldchangesII{This mechanism decreases} the 
number of errors induced \oldchangesI{by read operations to neighboring cells} 
by 72\%, which translates to a 16\% increase in NAND flash memory lifetime
(see Section~6.3 of our HPCA 2017 paper~\cite{cai.hpca17} for more detail).

\chVII{We conclude that, by} eliminating or reducing the 
probability of introducing errors during two-step programming, our 
solutions completely close or greatly reduce the exposure \oldchangesI{to} 
reliability \oldchangesII{and security} vulnerabilities.

%% file: sections/related.tex

\section{Related Work}
\label{sec:related}

To our knowledge, our HPCA 2017 paper~\cite{cai.hpca17} is the first to 
(1)~experimentally characterize both program interference and read
disturb errors that occur due to the two-step 
programming method commonly used in MLC NAND flash memory; 
(2)~\oldchangesIII{reveal} new \oldchangesII{reliability and} security vulnerabilities 
\oldchangesII{exposed} by two-step programming in \oldchangesII{flash memory}; and 
(3)~develop novel solutions to reduce these 
vulnerabilities. 
\oldchangesII{We briefly describe related works in the areas of
DRAM and NAND flash memory.}
\chVII{We note that a thorough survey of error mechanisms in NAND flash memory is
provided in our recent works~\cite{cai.procieee.arxiv17, cai.procieee17,
cai.bookchapter.arxiv17}.}

\subsection{Read Disturb Errors in DRAM}
\label{sec:related:dramreaddisturb}

Commodity DRAM
chips that are sold and used in the field today exhibit read
disturb errors~\cite{kim.isca14}, also called \emph{RowHammer}-induced errors~\cite{mutlu.date17}, 
which are \emph{conceptually} similar to the read disturb
errors found in NAND flash memory (see Section~\ref{sec:errors:readdisturb}).
Repeatedly accessing the same row in DRAM can cause
bit flips in data stored in adjacent DRAM rows. In order to
access data within DRAM, the row of cells corresponding
to the requested address must be \emph{activated} (i.e., opened for
read and write operations). This row must be \emph{precharged}
(i.e., closed) when another row in the same DRAM bank
needs to be activated. Through experimental studies on a
large number of real DRAM chips, we show that when a
DRAM row is activated and precharged repeatedly (i.e.,
\emph{hammered}) enough times within a DRAM refresh interval,
one or more bits in physically-adjacent DRAM rows can be
flipped to the wrong value~\cite{kim.isca14}.

\begin{figure}[h]
  \centering
  \includegraphics[width=0.8\columnwidth]{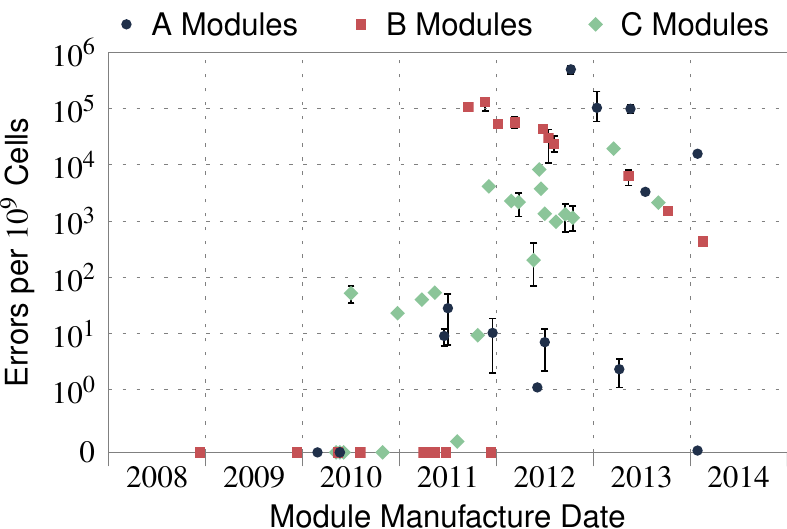}%
  \caption{RowHammer error rate vs.\ manufacturing dates of 129~DRAM
  modules we tested.  Reproduced from \cite{kim.isca14}.}%
  \label{fig:rowhammer-date}%
\end{figure}

\chVII{In our original RowHammer paper~\cite{kim.isca14}, we} tested 129~DRAM modules
manufactured by
three major manufacturers
(A, B, and C) between 2008 and 2014, using an FPGA-based experimental DRAM
testing infrastructure~\cite{hassan.hpca17} (more detail on our experimental
setup, along with a list of all modules and their characteristics, can be found 
in our original RowHammer paper~\cite{kim.isca14}).  
Figure~\ref{fig:rowhammer-date} shows the
rate of RowHammer errors that we found, with the 129~modules that we tested
categorized based on their manufacturing date.
We find that 110 of our tested modules exhibit RowHammer errors, with the
earliest such module dating back to 2010.  In particular, we find that
\emph{all} of the modules manufactured in 2012--2013 that we tested are
vulnerable to RowHammer. Like with many NAND flash memory error 
mechanisms, especially read disturb, RowHammer
is a recent phenomenon that especially affects DRAM chips manufactured with more advanced
manufacturing process technology generations\chVII{~\cite{mutlu.date17}.
The phenomenon is due to reliability problems caused by DRAM technology
scaling~\cite{mutlu.date17, mutlu.imw13, mutlu.superfri15, mutlu.memcon13}.}

\begin{figure}[h]
  \centering
  \includegraphics[width=0.8\columnwidth]{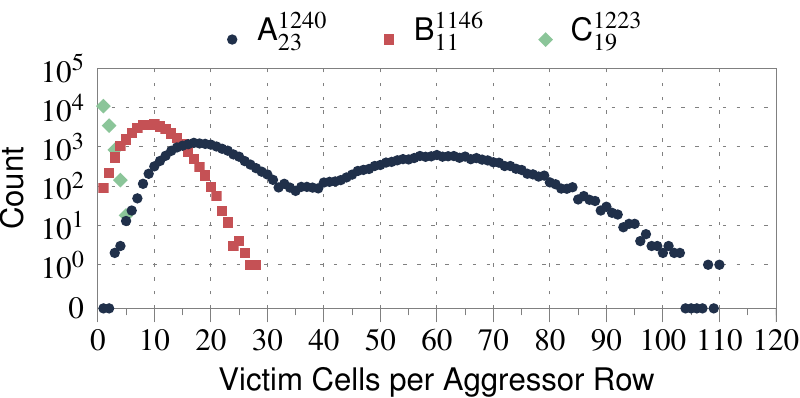}%
  \caption{Number of victim cells (i.e., number of bit errors) when an
  aggressor row is repeatedly activated, for three representative DRAM modules
  from three major manufacturers.  We label the modules in the format $X^{yyww}_n$, 
  where $X$ is the manufacturer (A, B, or C), $yyww$ is the manufacture year ($yy$) and
  week of the year ($ww$), and $n$ is the number of the selected module. 
  Reproduced from \cite{kim.isca14}.}%
  \label{fig:rowhammer-interference}%
\end{figure}

Figure~\ref{fig:rowhammer-interference} shows the distribution of
the number of rows (plotted in log scale on the y-axis) within a DRAM module that flip the
number of bits \chVII{shown} along the x-axis, as measured for example DRAM 
modules from three different DRAM manufacturers~\cite{kim.isca14}.
We make two observations from the figure.  First, the number of bits
flipped when we hammer a row (known as the \emph{aggressor row}) can vary
significantly within a module.  Second, each module has a different 
distribution of the number of rows.
Despite these differences, we find that this DRAM failure \chVII{mechanism}
affects more than 80\% of the DRAM chips we tested~\cite{kim.isca14}.
As indicated above, this read disturb error mechanism in
DRAM is popularly called RowHammer~\cite{mutlu.date17}.

Various recent works show that RowHammer can be
maliciously exploited by user-level software programs to
(1)~induce errors in existing DRAM modules~\cite{kim.isca14, mutlu.date17}
and (2)~launch attacks to compromise the security of various
systems~\cite{seaborn.blog15, mutlu.date17, seaborn.blackhat15,
gruss.dimva16, bosman.sp16, razavi.usenixsecurity16, vanderveen.ccs16,
burleson.dac16, xiao.usenixsecurity16, gruss.arxiv17}.
For example, by exploiting the RowHammer read disturb
mechanism, a user-level program can gain kernel-level
privileges on real laptop systems~\cite{seaborn.blog15, seaborn.blackhat15}, take over a
server vulnerable to RowHammer~\cite{gruss.dimva16}, take over a victim
virtual machine running on the same system~\cite{bosman.sp16}, and
take over a mobile device~\cite{vanderveen.ccs16}. Thus, the RowHammer
read disturb mechanism is a prime (and perhaps the
first) example of how a circuit-level failure mechanism in
DRAM can cause a practical and widespread system security
vulnerability.

Note that various solutions to RowHammer exist~\cite{kim.isca14, mutlu.date17, kim.thesis15},
but we do not discuss them in detail here.  Our recent work~\cite{mutlu.date17} 
provides a comprehensive overview.  A very promising proposal is to modify
either the memory controller or the DRAM chip such that it probabilistically
refreshes the physically-adjacent rows of a recently-activated row, with very
low probability.  This solution is called \emph{Probabilistic Adjacent Row
Activation} (PARA)~\cite{kim.isca14}.  Our prior work shows that this low-cost, low-complexity
solution, which does not require any storage overhead, greatly closes the
RowHammer vulnerability~\cite{kim.isca14}.

The RowHammer effect in DRAM worsens as the manufacturing
process scales down to smaller node sizes~\cite{kim.isca14, mutlu.date17}. 
More findings on RowHammer, along with extensive
experimental data from real DRAM devices, can be found in
our prior works~\cite{kim.isca14, mutlu.date17, kim.thesis15}.

\subsection{\chVII{Cell-to-Cell} Interference Errors in DRAM}
\label{sec:related:dramcelltocell}

Like NAND flash memory cells, DRAM cells are susceptible
to cell-to-cell interference. In DRAM, one important way
in which cell-to-cell interference exhibits itself is the data-dependent
retention behavior, where the retention time of
a DRAM cell is dependent on the values written to \emph{nearby}
DRAM cells\chVII{~\cite{mutlu.date17, liu.isca13, khan.sigmetrics14, khan.dsn16,
patel.isca17, khan.cal16, khan.micro17}}. This phenomenon
is called \emph{data pattern dependence} (DPD)~\cite{liu.isca13}. Data pattern
dependence in DRAM is similar to the data-dependent
nature of program interference that exists in NAND flash
memory (see Section~\ref{sec:errors:celltocell}). Within DRAM, data
\chVII{pattern} dependence
occurs as a result of parasitic capacitance coupling (between
DRAM cells). Due to this coupling, the amount of charge
stored in one cell's capacitor can inadvertently affect the
amount of charge stored in an adjacent cell's capacitor\chVII{~\cite{mutlu.date17, liu.isca13, khan.sigmetrics14, khan.dsn16, patel.isca17, khan.cal16, khan.micro17}}. 
As DRAM cells become smaller with
technology scaling, cell-to-cell interference worsens because
parasitic capacitance coupling between cells increases~\cite{liu.isca13, khan.sigmetrics14}. 
More findings on cell-to-cell interference and the
data-dependent nature of cell retention times in DRAM,
along with experimental data obtained from modern DRAM
chips, can be found in our prior works\chVII{~\cite{mutlu.date17, liu.isca13, khan.sigmetrics14, khan.dsn16, patel.isca17, khan.cal16, khan.micro17}}.

\subsection{Errors in Emerging Memory Technologies}
\label{sec:related:emerging}

Emerging
nonvolatile memories, such as \emph{phase-change memory} (PCM)~\cite{lee.isca09, qureshi.isca09, wong.procieee10, lee.ieeemicro10, zhou.isca09, lee.cacm10, yoon.taco14}, 
\emph{spin-transfer torque
magnetic RAM} (STT-RAM or STT-MRAM)~\cite{naeimi.itj13, kultursay.ispass13}, \emph{metal-oxide
resistive RAM} (RRAM)~\cite{wong.procieee12}, and \emph{memristors}~\cite{chua.tct71, strukov.nature08},
are expected to bridge the gap between DRAM and NAND-flash-memory-based SSDs, providing
DRAM-like access latency and energy, and at the same
time SSD-like large capacity and nonvolatility (and hence SSD-like
data persistence). 
While their underlying designs are different from DRAM and NAND flash memory,
these emerging memory technologies have been shown to exhibit similar types of
errors.
PCM-based devices are expected to have
a limited lifetime, as PCM can only \chVII{sustain a limited} number
of writes~\cite{lee.isca09, qureshi.isca09, wong.procieee10}, similar to the P/E cycling errors in
SSDs (though PCM's write endurance
is higher than that of SSDs\chVII{~\cite{lee.isca09}}). PCM suffers from (1)~\emph{resistance
drift}~\cite{wong.procieee10, pirovano.ted04, ielmini.ted07}, where the resistance used to represent the value
becomes higher over time (and eventually can introduce a bit error),
similar to how charge leakage in NAND flash memory and
DRAM lead to retention errors over time; and
(2)~\emph{write disturb}~\cite{jiang.dsn14}, where the heat generated during the programming of one
PCM cell dissipates into neighboring cells and can change the value that is
stored within the neighboring cells, similar in concept
to cell-to-cell program interference in NAND flash memory.
STT-RAM suffers from (1)~\emph{retention failures}, where the value
stored for a single bit (as the magnetic orientation of the layer that 
stores the bit) can flip over time; and (2)~\emph{read disturb} (a
conceptually different phenomenon
from the read disturb in DRAM and flash memory), where
reading a bit in STT-RAM can inadvertently induce a write to
that \chVII{\emph{same}} bit~\cite{naeimi.itj13}. 

Due to the nascent nature of emerging
nonvolatile memory technologies and the lack of availability of
large-capacity devices built with them, extensive and dependable
experimental studies have yet to be conducted on the reliability
of real PCM, STT-RAM, RRAM, and memristor chips.
However, we believe that error mechanisms conceptually or
abstractly similar to those for flash memory and DRAM are likely
to be prevalent in emerging technologies as well
(as supported by some recent studies~\cite{naeimi.itj13, jiang.dsn14, 
zhang.iccd12, khwa.isscc16, athmanathan.jetcas16, sills.vlsic15, sills.vlsit14}), 
albeit with different underlying mechanisms and error rates.

\subsection{Other Related Works}
\label{sec:related:other}

\paratitle{Memory Error Characterization and Understanding}
Prior works \oldchangesII{study} various types of NAND flash memory errors 
derived from circuit-level noise, such as 
data retention noise~\cite{cai.date12,cai.hpca15,cai.iccd12,cai.itj13, 
cai.procieee17, cai.procieee.arxiv17, cai.bookchapter.arxiv17, luo.hpca18,
mizoguchi.imw17, mielke.irps08}, 
read \oldchangesIII{disturb} noise~\cite{cai.dsn15, cai.procieee17,
cai.procieee.arxiv17, cai.bookchapter.arxiv17, papandreou.glsvlsi14, mielke.irps08}, 
cell-to-cell program interference noise{~\cite{cai.date12,cai.sigmetrics14,
cai.iccd13,cai.itj13}, and
P/E cycling noise\oldchangesIV{~\cite{cai.date13,cai.date12,parnell.globecom14,
luo.jsac16,cai.itj13, cai.procieee17, cai.procieee.arxiv17, cai.bookchapter.arxiv17, mielke.irps08}}.
Other prior works examine the aggregate effect of these errors on large sets
of SSDs that are deployed in the production data centers of
Facebook~\cite{meza.sigmetrics15}, Google~\cite{schroeder.fast16}, and Microsoft~\cite{narayanan.systor16}.
None of these works characterize how program interference and 
read disturb significantly increase errors within the unprogrammed or 
partially-programmed cells of an open block due to the vulnerabilities in 
two-step programming, nor \oldchangesII{do} they develop mechanisms that exploit 
or mitigate such errors.

A concurrent work by Papandreou et al.~\cite{papandreou.imw16} characterizes
the impact of read disturb on partially-programmed and unprogrammed cells in 
state-of-the-art MLC NAND flash memory.  The authors come to similar 
conclusions as we do about the impact of read disturb.  However, unlike our work,
they do not (1)~characterize the impact of cell-to-cell program interference on
partially-programmed cells, (2)~propose exploits that can take advantage of the
vulnerabilities in partially-programmed cells, or (3)~propose mechanisms that
mitigate or eliminate the vulnerabilities.

Similar to the characterization studies performed for NAND flash memory,
DRAM latency, reliability, and variation have been experimentally characterized
at both a small scale (e.g., hundreds of chips)~\chVII{\cite{liu.isca13, lee.hpca15,
chang.sigmetrics16, qureshi.dsn15, khan.sigmetrics14, lee.sigmetrics17, khan.dsn16,
patel.isca17, khan.micro17, khan.cal16, hassan.hpca17, kim.isca14,
chang.sigmetrics17, kim.hpca18, kim.thesis15, lee.thesis16, chang.thesis17}}
and a large scale (e.g., tens of thousands of chips)~\cite{meza.dsn15, schroeder.sigmetrics09, sridharan.asplos15, sridharan.sc13, hwang.asplos12}.

\paratitle{Program Interference Error Mitigation Mechanisms}
Prior works~\cite{cai.sigmetrics14,cai.iccd13} model the behavior of program interference, and 
\oldchangesII{propose mechanisms that estimate} the optimal read reference 
voltage once interference has occurred. 
These works minimize program interference errors \oldchangesII{only} for 
\emph{fully-programmed} wordlines, \oldchangesII{by modeling the change in the
threshold voltage distribution as a result of the interference.
These models are fitted to the distributions of wordlines after \oldchangesIII{\emph{both}} the 
LSB and MSB pages are programmed, and are unable to determine and \oldchangesIII{mitigate}
the shift that occurs for wordlines that \oldchangesIII{are \emph{partially 
programmed}}.}
In contrast, we propose mechanisms that specifically address the program 
interference resulting from two-step programming, and \oldchangesIII{reduce the number 
of errors induced on} LSB pages \oldchangesIII{in} 
\oldchangesIV{\emph{both}} partially-programmed and unprogrammed wordlines.

\paratitle{Read Disturb Error Mitigation Mechanisms}
One patent~\cite{frost.patent10} proposes \oldchangesII{a mechanism that uses} counters to monitor 
the total number of reads to each block.  \oldchangesII{Once} a block's counter 
exceeds a threshold, \oldchangesII{the mechanism remaps and rewrites all of the 
valid pages within the block} to remove the accumulated read disturb 
errors~\cite{frost.patent10}.
Another patent~\cite{schushan.patent14} \oldchangesII{proposes to monitor} the MSB page 
error rate to ensure that it does not exceed the \oldchangesII{ECC error correction
capability}, to avoid data loss. 
\oldchangesII{Both of these} mechanisms monitor pages \oldchangesII{\emph{only}} from 
\emph{fully-programmed wordlines}. Unfortunately, as we observed, 
LSB pages \oldchangesIII{in} partially-programmed and 
unprogrammed \oldchangesII{wordlines} are twice as susceptible to read disturb 
\oldchangesII{as pages \oldchangesIII{in} fully-programmed wordlines}
(see Section~\ref{sec:errors:readdisturb}). If only the MSB page 
error rate is monitored, read \oldchangesII{disturb} may be detected too late to correct 
some of the LSB pages. 

Our earlier work~\cite{cai.dsn15} dynamically changes \oldchangesII{the} pass-through voltage
for each block to reduce the impact of read disturb. As a 
single voltage is applied to the whole block, this mechanism does \emph{not} 
help significantly with the LSB pages \oldchangesIII{in} partially-programmed and 
unprogrammed wordlines. In contrast, our read disturb mitigation technique 
(see Section~\ref{sec:mech}) specifically targets these LSB pages by applying multiple
different pass-through voltages in an open block, optimized to \oldchangesII{the} 
different programmed states of each wordline, to reduce read disturb errors.

Other prior works~\cite{ha.apsys13, ha.tcad16, kim.patent12} propose to use
\emph{read reclaim} to mitigate read disturb errors.
The key
idea of read reclaim is to remap the data in a block to a new
flash block, if the block has experienced a high number of
reads~\cite{ha.apsys13, ha.tcad16, kim.patent12}.
Read reclaim is similar to the remapping-based refresh 
mechanism~\cite{cai.iccd12, cai.itj13, luo.msst15, mohan.tr12, pan.hpca12}
employed by many modern SSDs to mitigate data retention
errors\chVII{~\cite{cai.procieee.arxiv17, cai.procieee17, cai.bookchapter.arxiv17}}.
Read reclaim can remap the contents of a wordline only after the wordline is
fully programmed, and does \chVII{\emph{not}} mitigate the impact of read disturb on
partially-programmed or unprogrammed wordlines.

\paratitle{Using Flash Memory for Security Applications}
Some \oldchangesII{prior} works studied how \oldchangesII{flash memory} can be used to enhance the 
security of applications. 
One work~\cite{wang.sp13} uses \oldchangesII{flash memory} as a secure channel to hide information, such as a 
secure key. 
Other works~\cite{wang.sp12,xu.imw14} use \oldchangesII{flash memory} to generate random numbers and 
digital fingerprints. None of these works study 
vulnerabilities that exist within the \oldchangesII{flash memory}.

Based on our HPCA 2017 paper~\cite{cai.hpca17}, recent work~\cite{kurmus.woot17}
demonstrates how an attack can be performed on a real SSD using our program 
interference based exploit (see Section~\ref{sec:exploits:celltocell}).  
The authors use our exploit to
perform a file system level attack on a Linux machine, using the attack to 
gain root privileges.

\paratitle{\oldchangesII{Two-Step vs. One-Shot} Programming}
One-shot programming shifts flash cells directly from the erased state to their 
final target state in a single step. For smaller transistors with less distance 
between neighboring flash cells, such as those in sub-40nm planar (i.e., 2D) NAND flash 
memory, two-step programming \oldchangesII{has} replaced \oldchangesIII{one-shot} programming 
to alleviate the coupling capacitance resulting from cell-to-cell program 
interference~\cite{park.jssc08}. 3D NAND \oldchangesIII{flash memory} 
currently uses one-shot programming~\cite{parnell.fms16, yoon.fms17, parnell.fms17}, 
as 3D NAND flash memory chips use \oldchangesII{larger process technology nodes} 
(i.e., 30--50~nm)~\cite{yoon.fms15, samsung.whitepaper14} and employ
charge trap transistors~\cite{park.jssc15, katsumata.vlsit09, wegener.iedm67, eitan.patent98, komori.iedm08, tanaka.vlsit07, jang.vlsit09} for flash cells, as opposed to the floating-gate transistors
used in planar NAND flash memory. However, 
once the number of \oldchangesIII{3D-stacked layers} reaches its upper limit~\cite{lapedus.semieng16, lee.eetimes17}, 3D NAND flash memory is expected to scale to 
smaller transistors~\cite{yoon.fms15}, and we expect that the increased program interference will 
again require partial programming (just as it happened for planar NAND flash memory
\oldchangesII{in the past\oldchangesIII{~\cite{park.jssc08, kim.irps10}}}). 
\chVII{More detail on 3D NAND flash memory is provided in a recent survey
article~\cite{cai.bookchapter.arxiv17}.}

%% file: sections/impact.tex

\section{Long-Term Impact}

As we discuss in Section~\ref{sec:related}, our \chVII{HPCA 2017} paper~\cite{cai.hpca17} 
makes several novel contributions on characterizing, exploiting, and mitigating
vulnerabilities in the two-step programming algorithm used in state-of-the-art
MLC NAND flash memory.
We believe that these contributions \chVII{are likely to} have a significant impact on
academic research and industry.

\subsection{Exposing the Existence of Errors}

NAND flash manufacturers use two-step programming widely in their contemporary
MLC NAND flash devices.  Prior to our HPCA 2017 paper~\cite{cai.hpca17} and
concurrent work by Papandreou et al.~\cite{papandreou.imw16}, there was no publicly-available knowledge 
about how two-step programming introduced new error sources that did \chVII{\emph{not}}
exist in the prior one-shot programming approach.
Using real off-the-shelf contemporary NAND flash memory chips,
our HPCA 2017 paper exposes the fact that fundamental limitations of the two-step programming
method introduce program errors that reduce the lifetime of SSDs available on
the market today.

Through a rigorous characterization, our HPCA 2017 paper~\cite{cai.hpca17} analyzes two major sources of these errors, 
program interference and read disturb, demonstrating how they can corrupt 
data stored in a partially-programmed flash cell.
While prior works have addressed both program interference
(e.g.~\cite{cai.iccd13,park.jssc08,dong.tcas10,lee.edl02})
and read disturb (e.g.,~\cite{cai.dsn15,ha.apsys13,frost.patent10,schushan.patent14})
errors in the past, we find that none of these existing solutions \chVII{are able} to
protect the vulnerable partially-programmed pages produced during
two-step programming.
We expect that by exposing these errors and the unique vulnerabilities of
partially-programmed cells, our work will (1)~provide NAND flash memory 
manufacturers and the academic community with
significant insight into the problem; (2)~foster the development of
new solutions that can reduce or eliminate this vulnerability; and
(3)~inspire others to search for other \chVII{reliability and security} vulnerabilities that exist in
NAND flash memory.

\subsection{Security Implications for Flash Memory}

Our HPCA 2017 paper~\cite{cai.hpca17} proposes two sketches of new potential security exploits based on 
errors arising from two-step programming. Malicious applications can 
\oldchangesII{be developed to}
use these (or other similar) exploits to corrupt data belonging to other applications.
For example, our paper has already enabled the development and demonstration
of a file system based attack by IBM security researchers~\cite{kurmus.woot17}.
In that work, the researchers built upon our program interference based
exploits to show how to use the file system to acquire root privileges on a
real machine.  The work confirms that our exploit sketches are \chVII{likely} viable on
a real system, and that the threat of maliciously exploiting vulnerabilities in two-step
programming is real (and needs to be addressed).

As was the case for \chVII{RowHammer} attacks in DRAM \chVII{(see Section~\ref{sec:related:dramreaddisturb})}, our findings have already
generated significant interest and concern in the broader \chVII{technology} community
(e.g.,~\cite{burton.inquirer,
hruska.extremetech, cimpanu.bleepingcomputer, chirgwin.register}).
The reason behind the broader impact of our work is that many existing drives in
the field today can be attacked.  After IBM researchers demonstrated the 
ability to perform such attacks on a real system~\cite{kurmus.woot17}, there has been further
interest in NAND flash memory attacks (e.g.,~\cite{mimoso.threatpost, aldershoff.myce}).

We hope and expect that other researchers will take our cue and begin to
investigate how other reliability issues in NAND flash memory can be 
exploited by applications to perform malicious attacks.  We believe that
this is a new area of research that will grow in importance as SSDs \chVII{and flash memory} become
even more widely used.

\subsection{Eliminating Program Error Attacks}

Our HPCA 2017 paper~\cite{cai.hpca17} proposes \oldchangesII{three} solutions that either eliminate or mitigate 
vulnerabilities \oldchangesII{to program interference and read disturb}
during two-step programming. 
We intentionally design all three of our solutions to be low overhead and
easily implementable in commercial SSDs.  One of
our three solutions completely eliminates the vulnerabilities, albeit with a small
increase in \oldchangesII{flash} programming latency.
We expect our work to have a direct
impact on the NAND flash memory industry, as manufacturers will likely
incorporate solutions such as the ones we propose to mitigate or
eliminate these vulnerabilities in their new SSDs.
We also expect manufacturers and researchers to explore new mechanisms,
inspired by our work and by our solutions, that can eliminate these or
other vulnerabilities and exploits due to NAND flash memory reliability errors.

%% file: sections/conclusion.tex

\section{Conclusion}
\label{sec:conclusion}

Our HPCA 2017 paper~\cite{cai.hpca17} shows that the two-step programming mechanism commonly employed in modern MLC NAND flash memory chips opens up new vulnerabilities to errors, based on an experimental characterization of modern 1X-nm MLC NAND flash chips. We show that the root cause of these vulnerabilities is the fact that when a partially-programmed cell \oldchangesIII{is set to} an intermediate threshold voltage, it is much more susceptible to both cell-to-cell program interference and read \oldchangesIII{disturb}. We demonstrate that (1)~these vulnerabilities lead to errors that reduce the overall reliability of flash memory, and (2)~attackers can \oldchangesIII{potentially} exploit these vulnerabilities to maliciously corrupt data belonging to other programs. Based on our experimental observations and the resulting understanding, we propose \oldchangesIII{three} new mechanisms that can remove or mitigate these vulnerabilities, by eliminating or reducing the errors introduced as a result of the two-step programming method. Our experimental evaluation shows that our new mechanisms are effective: they can either eliminate the vulnerabilities with modest/low latency overhead, or drastically reduce the vulnerabilities and reduce errors with negligible latency or storage overhead. We hope that the vulnerabilities \changes{we} analyzed and exposed in this work\oldchangesIII{, along with the experimental data \changes{we} provided,} open up new avenues for mitigation as well as for exposure of other potential vulnerabilities due to internal flash memory operation.